\documentclass[preprint,tightenlines,superscriptaddress]{revtex4-2}

\usepackage[utf8]{inputenc}
\usepackage{amsmath,amssymb}
\usepackage{braket}
\usepackage[colorlinks=true,linkcolor=blue,urlcolor=blue,citecolor=blue,anchorcolor=blue]{hyperref}
\usepackage[capitalise]{cleveref}
\Crefname{section}{Sec.}{Secs.}
\crefrangelabelformat{equation}{(#3#1#4--#5#2#6)}
\usepackage{tikz}
\usepackage{multirow}
\usepackage{amsfonts}
\usepackage{ulem}
\usepackage{diagbox}
\usepackage{subfigure}

\newcommand{\ie}{{i.e.}}
\newcommand{\eg}{{e.g.}}

\newcommand{\tr}{\operatorname{tr}}

\begin{document}
\title{Inverse scattering problem with a bare state}
\newcommand{\UCAS}{School of Physical Sciences, University of Chinese Academy of Sciences (UCAS), Beijing 100049, China}
\author{Yan Li}
\affiliation{\UCAS}
\author{Jia-Jun Wu}
\affiliation{\UCAS}

\begin{abstract}
    In hadron physics, molecular-like multihadron states can interact with compact multiquark states.
    The latter are modeled as bare states in the Hilbert space of a potential model.
    In this work, we study several potential models relevant to the bare state, and solve their inverse scattering problems.
    The first model, called ``cc'', is a separable potential model. 
    We show that it can approximate (single-channel short-range) $S$-wave near-threshold physics with an error of $\mathcal{O}(\beta^3/M_V^3)$, where $\beta$ sets the maximum momentum of the near-threshold region and $M_V$ is the typical scale of the potential. 
    The second model, called ``bc'', serves as the bare-state-dominance approximation, where interaction between continuum states is ignored.
    Under this model, even though the bare state is always crucial for a bound state's generation, a shallow bound state naturally tends to have a small bare-state proportion. 
    %
    Therefore, we need other quantities to quantify the importance of the bare state.
    The last model, called ``bcc'', is a combination of the first two models. 
    This model not only serves as a correction to the bare-state-dominance approximation, but can also be used to understand the interplay between quark and hadron degrees of freedom. 
    This model naturally leads to the presence of a Castillejo-Dalitz-Dyson (CDD) zero. 
    We consider the energy decomposition of a bound state. 
    The potential ratio of the bare-continuum interaction to the continuum self-interaction is proposed to understand how the bound state is generated.
    Model independently, an inequality for the potential ratio is derived.
    Based on the model ``bcc'', the CDD zero can be used to estimate the potential ratio.
    Finally, we apply these studies to the deuteron, $\rho$ meson, and $D_{s0}^*(2317)$, and analyze their properties.
\end{abstract}

\maketitle
\newpage

\section{Introduction} 
Quantum chromodynamics (QCD) is the well-established theory of the strong interaction.
It describes not only how quarks form hadrons, but also how hadrons interact with each other.
However, it is difficult to analytically understand these processes within QCD because of its nonperturbative nature.
Therefore potential models are still a useful tool in hadron physics.

On the quark level, there exists a confining potential between quarks, and hadrons are bound states in the system.
With a suitably designed potential, the model can have good predictive power.
For example, the well-known Godfrey-Isgur model \cite{Godfrey:1985xj} can well describe the mass spectrum of mesons. 
On the hadron level, hadrons interact with each other within a short-range potential.
Unlike the quark-level case where bound states can appear both below and above the threshold and no scattering can happen, the hadron-level case can only have bound states below the threshold and scattering happens above the threshold.
Various potential models have been used to study the binding and scattering between hadrons.
Examples include Hamiltonian effective field theory \cite{Hall:2013qba,Wu:2014vma,Hall:2014uca,Liu:2015ktc,Liu:2016wxq,Liu:2016uzk,Wu:2016ixr,Wu:2017qve,Li:2019qvh,Abell:2021awi,Li:2021mob,Yang:2021tvc} where a potential is parametrized respecting chiral effective field theory, the HAL QCD method \cite{Aoki:2020bew,Ishii:2006ec,HALQCD:2012aa} where a potential is reconstructed from the lattice simulated Nambu-Bethe-Salpeter wave function, and many potential models for the nuclear force like those in Refs.~\cite{Stoks:1994wp,Wiringa:1994wb,Machleidt:2000ge}.

The direct scattering problem starts with a potential, and solves scattering equations to get scattering observables.
The inverse scattering problem tries to reconstruct the potential from scattering observables.
Most studies in this field were done decades ago.
We refer interested readers to Ref.~\cite{Chadan:1989Inverse}.
There are also inverse scattering problems concerning the off-shell $T$ matrix instead of the potential (see, \eg, Ref.~\cite{Sauer:2019ive}), which is not the focus of the current work.

Among those studies, however, little attention has been paid to the bare state.
In hadron physics, compact multiquark states are modeled as bare states, and interact with molecular-like multihadron states.
For example, in the quark model, the $\rho$ meson is a stable bound state of a quark-antiquark pair, while in the real world it can decay to two pions.
In the potential model, in addition to the two-pion basis states, one can include the quark-antiquark pair as a bare state in the Hilbert space, and also include the interaction between them.
Other examples include the $\Lambda(1405)$ in the $\bar{K}N$-$(uds)$ system \cite{Hall:2014uca}, and the $D_{s0}^*(2317)$ in the $DK$-$(c\bar{s})$ system \cite{Yang:2021tvc}.
References \cite{Zhou:2020moj,Zhou:2020vnz} also used the bare state to explain the two-pole structure appearing in many systems.
As there are many such kinds of systems in hadron physics, the inverse scattering problem with a bare state is worth exploring, and it is the main purpose of this work.

Past works on the inverse scattering problem mostly focus on the reconstruction of the potential.
Though in principle one can get any quantity from the potential, it is still interesting to consider using observables to directly construct quantities that are not directly observed, \eg, the wave function of the bound state.
In addition, after including the bare state, many new attractive quantities appear, including the bare-state mass, the bare-state proportion in the bound state, and the bare-state distribution in the eigenmodes of the full Hamiltonian.
One of the central topics in hadron physics is to understand the structure of hadrons. 
For the examples introduced above, we want to know if $\Lambda(1405)$ is mostly a $\bar{K}N$ molecular state or a $(uds)$ compact baryon state, and if $D_{s0}^*(2317)$ is a $DK$ molecular state or a $(c\bar{s})$ compact meson state.
These questions can be summarized in the potential model as what is the probability of finding a bound state in a bare state.

Although he did not start from the inverse scattering problem, Weinberg \cite{Weinberg:1965zz} found approximate relations between the compositeness of a shallow bound state with the scattering length and effective range.
Weinberg's relations and their extensions \cite{Baru:2003qq,Gamermann:2009uq,Baru:2010ww,Hanhart:2011jz,Hyodo:2011qc,Aceti:2012dd,Hyodo:2013iga,Hyodo:2013nka,Sekihara:2014kya,Hanhart:2014ssa,Guo:2015daa,Sekihara:2015gvw,Kamiya:2015aea,Xiao:2016dsx,Xiao:2016wbs,Kang:2016ezb,Sekihara:2016xnq,Kamiya:2016oao,Guo:2016wpy,Oller:2017alp,Kamiya:2017hni,Bruns:2019xgo,Matuschek:2020gqe,Li:2021cue,Albaladejo:2022sux,Bruns:2022hmb,Song:2022yvz} have been widely used to detect the structure of many hadrons (see Ref.~\cite{Guo:2017jvc} for a review).
One of the extensions \cite{Li:2021cue} by us and our collaborators makes use of techniques from the inverse scattering problem.

Though potential models still prevail in current studies of hadron physics, the inverse problem receives only a little attention.
Perhaps the main reason is that the potential can hardly be uniquely determined by observables.
However, the success of Weinberg's exploration mentioned before shows an opportunity that one can still find inverse scattering relations under certain approximations.
In this work, we take this opportunity by discussing how some potential models can serve as approximations, and study their inverse scattering problems.

In \cref{sec:fd} we study the Fredholm determinant, which is useful for discussions of inverse scattering problems.
In \cref{sec:toy} several potential models are discussed.
In \cref{sec:rws} those models are applied to analyze real-world systems, including the deuteron, $\rho$ meson, and $D_{s0}^*(2317)$.

\section{Fredholm determinant}\label{sec:fd}

In this paper, we focus on single-continuum-channel systems with or without a single bare state. 
In the current section, however, systems with an arbitrary number of bare states are also included.
For a partial-wave-projected Hamiltonian $H=H_0+V$, the Fredholm determinant is defined as
\begin{align}\label{eq:Ddef}
    D(W) := \det \left( 1 - \frac{1}{W-H_0}V \right) = \det\left[\frac{1}{W-H_0}(W-H)\right] \,.
\end{align}
In various cases, including the local potential \cite{Jost:1951zz,Newton:1972wd}, the nonlocal potential \cite{Warke:1971sp,Singh:1971nd}, and the potential with bare states generated from confined channels \cite{Dashen:1976nn,Dashen:1976cf,Vidal:1992uj,Vidal:1992ye}, when $E$ is above the threshold $E_{\text{th}}$, $D(E+i\varepsilon)$ satisfies
\footnote{
The Fredholm determinant may diverge in certain cases. 
Ref.~\cite{Weinberg:1963zza}, for instance, showed that $D(W)$ diverges for the Coulomb potential due to its long-range nature. 
Ref.~\cite{Weinberg:1963zza} also showed it, however, that the Hulth\'{e}n potential whose $D(W)$ is finite, can be used to approximate the Coulomb potential. 
In this work, we just assume that $D(W)$ is made to be well-defined.
}
\begin{align}\label{eq:Ddelta}
    \arg D(E+i\varepsilon) = - \delta(E) \mod \pi \,,
\end{align}
where $\delta$ is the scattering phase shift.
For further convenience, we also list some useful alternatives to \cref{eq:Ddelta}:
\begin{align}
    e^{i\delta(E)}\sin\delta(E) &= -\frac{\text{Im}\,D(E+i\varepsilon)}{D(E+i\varepsilon)} \,,\label{eq:esin}\\
    \tan \delta(E) &= -\frac{\text{Im}\,D(E+i\varepsilon)}{\text{Re}\,D(E+i\varepsilon)} \,.\label{eq:tan}
\end{align}

From \cref{eq:Ddef}, it is easy to see that $D(W)$ has zeros residing on the bound states' energies $E_{B_i}$, poles on the bare states' energies $m_{b_i}$, and a branch cut starting from the threshold $E_\text{th}$ of the only continuum channel to positive infinity.
The zeros and singularities of $D(W)$ prevent its phase from being continuous, so, after introducing
\begin{align}\label{eq:cw}
    C(W) := \frac{\prod_{i=1}^{N_B} \frac{W-E_{B_i}}{W-E_\text{th}}}{\prod_{i=1}^{N_b} \frac{W-m_{b_i}}{W-E_\text{th}}}
\end{align}
(where $N_b$ and $N_B$ are the numbers of bare states and bound states respectively), $D(W)/C(W)$ will then have a continuous phase, except for possibly at the threshold.
To completely specify the phase convention of $D(W)/C(W)$, we set its phase at complex infinity to be zero:
\begin{align}
    \arg  \frac{D(\infty)}{C(\infty)} =0\,.
\end{align}
Because $\text{Im }D(E)/C(E)=0$ below threshold, $\arg D(E)/C(E)=\pi n$ for some integer $n$ for all $E<E_{\text{th}}$. 
Then $\arg D(E)/C(E)=\arg D(\infty)/C(\infty) =0$ for all $E<E_{\text{th}}$.

In this work, we also specify the phase convention of the scattering phase shift as
\begin{align}\label{eq:ddelta}
    \delta(E):=-\arg \frac{D(E+i\varepsilon)}{C(E+i\varepsilon)}\,.
\end{align}
From \cref{eq:cw} we know that $D(W)/C(W)\sim (W-E_{\text{th}})^{N_B-N_b}$ around the threshold, and thus we have 
\begin{align}
    \delta(\infty)-\delta(E_\text{th}) &= -\left[\arg\frac{D(\infty)}{C(\infty)} -\arg\frac{D(E_\text{th}+\varepsilon_0+i\varepsilon)}{C(E_\text{th}+\varepsilon_0+i\varepsilon)}\right] \nonumber\\
    &= -\left[\arg\frac{D(E_\text{th}-\varepsilon_0)}{C(E_\text{th}-\varepsilon_0)} -\arg\frac{D(E_\text{th}+\varepsilon_0+i\varepsilon)}{C(E_\text{th}+\varepsilon_0+i\varepsilon)}\right] \nonumber\\
    &=-(N_B-N_b) \left[\arg(-\varepsilon_0) - \arg(\varepsilon_0+i\varepsilon)\right] = -\pi(N_B-N_b)
\end{align}
where we let $\varepsilon \ll \varepsilon_0$, although they are both infinitesimals.
The above relation between $\delta(E_\text{th})$ and $\delta(\infty)$ is actually a part of the generalized Levinson theorem derived in Ref.~\cite{Vidal:1992uj} (Theorem III) for multichannel scattering with confining potentials.
In our convention, we also have $\delta(\infty)=0$, so $\delta(E_\text{th})=\pi(N_B-N_b)$.
Theoretically, the number of bound states and the phase shift are observables, so the Levinson theorem will tell us the number of bare states. 
In reality, the phase shift at infinity can hardly be recognized as an observable, but the Levinson theorem can still provide a qualitative estimate of the number of bare states.

$D(W)/C(W)$ is a real analytic function, allowing a standard dispersive analysis that gives
\begin{align}
    \ln \frac{D(W)}{C(W)} &= - \frac{1}{\pi} \int_{E_\text{th}}^{\infty} dE \frac{\delta(E)}{E-W} \,,
\end{align}
where we have assumed 
\begin{align}
    \lim_{|W|\to\infty} D(W) = 1 \,.
\end{align}
Then the dispersive representation of $D(W)$ reads
\begin{align}\label{eq:dispDW}
    D(W) = \exp\left(- \frac{1}{\pi} \int dE \frac{\delta(E)}{E-W} \right)C(W) \,.
\end{align}

In addition, by noting that
\begin{align}
    \lim_{|W|\to\infty} W\ln D(W) &= \lim_{|W|\to\infty} W\ln \det\left(1-\frac{1}{W-H_0}V\right) \nonumber\\
    &=\lim_{|W|\to\infty} \operatorname{tr} \ln \left(1-\frac{1}{W-H_0}V\right)^W \nonumber\\
    &=\operatorname{tr}\ln\exp (-V) = -\operatorname{tr}V \,,
\end{align}
and
\begin{align}
    \lim_{|W|\to\infty} W\ln D(W) &= \lim_{|W|\to\infty} W\ln \left[\exp\left(-\frac{1}{\pi}\int dE \frac{\delta(E)}{E-W}\right) C(W)\right] \nonumber\\
    &= \lim_{|W|\to\infty} \left[-\frac{W}{\pi}\int dE \frac{\delta(E)}{E-W} +\ln C(W)^W \right] \nonumber\\
    &=  \frac{1}{\pi}\int dE \,\delta(E) - \sum_{i=1}^{N_B} (E_{B_i}-E_\text{th}) + \sum_{i=1}^{N_b} (m_{b_i}-E_\text{th})\,,
\end{align}
one gets a trace formula:
\begin{align}\label{eq:trf}
    \tr V  = \sum_{i=1}^{N_B} (E_{B_i}-E_\text{th}) - \sum_{i=1}^{N_b} (m_{b_i}-E_\text{th}) - \frac{1}{\pi}\int dE \,\delta(E) \,.
\end{align}


\section{Several potential models}\label{sec:toy}

In this section, we study the inverse scattering problem for several potential models.
We label the first one as ``cc'', which has a single continuum channel and no bare states, and it is also known as the separable potential model elsewhere. 
Its Hamiltonian reads
\begin{align}\label{eq:Hcc}
    H &= H_0 + V = \int \frac{p^2 dp}{(2\pi)^3} h_p \ket{p}\bra{p} + \lambda_c\ket{f}\bra{f} \,, 
\end{align}
where $\ket{p}$ are the noninteracting momentum-space continuum states normalized as
\begin{align}
    \braket{p|k}=\frac{(2\pi)^3}{p^2}\delta(p-k) \,,
\end{align}
$h_p$ is the noninteracting energy, and
\begin{align}
    \ket{f} = \int\frac{p^2dp}{(2\pi)^3} f(p) \ket{p} \,.
\end{align}
For the $l$th partial wave, we have $f(p) \to p^l$ when $p\to0$. 
Without loss of generality, $f(p)$ is assumed to be real.

The second model, called ``bc'', includes a continuum channel and a bare state. 
It has the following Hamiltonian:
\begin{align}\label{eq:Hbc}
    H &= H_0 + V = \left( m \ket{b}\bra{b} + \int \frac{p^2 dp}{(2\pi)^3} h_p \ket{p}\bra{p} \right) +  \lambda_b \left( \ket{b}\bra{f} + \ket{f}\bra{b}\right) \,. 
\end{align}
This model ignores the interaction between the continuum states.
It is also known as the Friedrichs-Lee model \cite{Friedrichs:1948perturbation,Lee:1954iq}.

The last model, called ``bcc'', still has a continuum channel and a bare state. 
It is a combination of the first two models:
\begin{align}\label{eq:Hbcc}
    H &= H_0 + V = \left( m \ket{b}\bra{b} + \int \frac{p^2 dp}{(2\pi)^3} h_p \ket{p}\bra{p} \right) +  \Big[ \lambda_b \left( \ket{b}\bra{f} + \ket{f}\bra{b} \right) + \lambda_c \ket{f}\bra{f} \Big] \,.
\end{align}

\subsection{Model ``cc''}\label{sec:cc}
The model ``cc'', or the separable potential model, has been widely used mostly because it is easily soluble.
Here we discuss how it can provide a good approximation for the (single-channel short-range) $S$-wave near-threshold physics.
Let us assume a real-world system described by a Hamiltonian with a specific potential $V^{\text{(rw)}}(\vec{p},\vec{k})$.
We label the typical momentum scale of $V^{\text{(rw)}}(p,k)$ as $M_V$.
From dimensional analysis, we should have $V^{\text{(rw)}}(p,k)=\mathcal{O}(M_V^{-2})$. 
For example, the well-known Yukawa potential originating from one-boson exchange reads
\begin{align}\label{eq:yukawa}
    V(\vec{p},\vec{k}) \propto \frac{1}{(\vec{p}-\vec{k})^2+M_V^2} = \mathcal{O}(M_V^{-2}) \,,
\end{align}
which after $S$-wave projection should still be of $\mathcal{O}(M_V^{-2})$.
Now we want to study how the model ``cc'' can approximate the near-threshold regime of this system where $p,k$ are smaller than a momentum $\beta$.

The potential $V^{\text{(rw)}}(\vec{p},\vec{k})$ is a function of $\vec{p}$ and $\vec{k}$.
The rotational invariants built from them are $\vec{p}^2$, $\vec{k}^2$, and $\vec{p}\cdot\vec{k}=p\,k\,\cos\theta$.
Then in the low-momentum region, the potential can be expanded as a power series in terms of these invariants:
\begin{align}
    V^{\text{(rw)}}(\vec{p},\vec{k})=c_0 +c_1\,(\vec{p}^2 + \vec{k}^2) + c_2\,\vec{p}\cdot\vec{k} + \cdots \,.
\end{align}
After partial-wave projection, the $\vec{p}\cdot\vec{k}$ term disappears in the $S$ wave.
Then in the low-momentum region, the corresponding $S$-wave-projected potential $V^{\text{(rw)}}(p,k)$ can be expanded as
\begin{align}\label{eq:exvpk}
    V^{\text{(rw)}}(p,k) = v_0 + v_1\,(p^2 + k^2) + \cdots \,.
\end{align}
In the meantime, we can also expand the ``cc'' potential as
\begin{align}\label{eq:exfpk}
    \lambda_c\,f(p)\,f(k) = \lambda_c\,(f_0 + f_1\,p^2 + \cdots)(f_0 + f_1\,k^2 + \cdots) = \lambda_c\,f_0^2 + \lambda_c\,f_0\,f_1(p^2+k^2) + \cdots \,.
\end{align}
So by setting $\lambda_c\,f_0^2=v_0$ and $\lambda\,f_0\,f_1=v_1$, the ``cc'' potential can reproduce any real-world potential up to the second order of momenta.
So the relative error of approximating $V^{\text{(rw)}}(p<\beta,k<\beta)$ is of a higher order: $\mathcal{O}\left(\beta^4/M_V^4 \right)$.
However, one cannot just conclude that the final error is the same, because $V^{\text{(rw)}}(p,k)$ of higher momenta can also couple to the low-momentum physics.
For example, the second term in the Born series:
\begin{align}
    \int \frac{q^2dq}{(2\pi)^3} \frac{V^{\text{(rw)}}(p,q)V^{\text{(rw)}}(q,k)}{E-h_{q}} \,,
\end{align} 
can receive a contribution of $\mathcal{O}\left[ \frac{1}{M_V^2}\frac{\mu}{M_V} \right]$ in the $q\sim M_V$ region of integration (we assume $\mu > M_V$ so that the nonrelativistic approximation $h_{M_V}\approx M_V^2/2\mu$ works), which is even larger than the first term in the Born series: $V^{\text{(rw)}}(p,k)=\mathcal{O}(M_V^{-2})$.
In \cref{app:assep}, we provide an error analysis, taking care of the high-momentum region, where we show that the model ``cc'' can typically approximate the near-threshold physics of a system with a relative error of $\mathcal{O}\left( \beta^3/M_V^3  \right)$, providing the absence of a deeply bound state.

Now we come to study the model ``cc''.
This model was already studied decades ago in Refs.~\cite{Gourdin:1957Interaction,Gourdin:1958Exact}.
We review and extend their discussions in our notations for completeness.

We first work out the Fredholm determinant:
\begin{align}\label{eq:Dcc}
    D(W)=1 - \lambda_c \bra{f}\frac{1}{W-H_0}\ket{f} \equiv 1-\lambda_c\,F(W) \,.
\end{align}
Its imaginary part satisfies
\begin{align}\label{eq:DImcc}
    \mathrm{Im}\,D(h_p+i\varepsilon)/\lambda_c = \pi \bra{f}\delta_D(h_p-H_0)\ket{f} = \frac{\pi\,p^2}{(2\pi)^3 h'_p} f^2(p) \geq 0 \,,
\end{align}
where $\delta_D$ is the Dirac delta function, and $h'_p=\frac{d}{dp}h_p$.
Using \cref{eq:Dcc,eq:DImcc}, one finds the following [note that $\delta(\infty)=0$ by convention]:
\begin{itemize}
    \item $\lambda_c>0$ (repulsive, no bound states): $\delta \in [-\pi,0]$ and $\delta(E_\text{th})=0$.
    \item $F(E_\text{th})^{-1}<\lambda_c<0$ (attractive, no bound states): $\delta \in [0,\pi]$ and $\delta(E_\text{th})=0$.
    \item $\lambda_c<F(E_\text{th})^{-1}$ (attractive, one bound state): $\delta \in [0,\pi]$ and $\delta(E_\text{th})=\pi$.
\end{itemize}

The Fredholm determinant has the dispersive representation \cref{eq:dispDW} with
\begin{align}\label{eq:Ccc}
    C(W) = \begin{cases}
        \;\;\;\;1\,,&N_B=0\,,\\
        \frac{W-E_B}{W-E_\text{th}}\,,&N_B=1\,.\\
       \end{cases} \,.
\end{align}
Taking the imaginary part of \cref{eq:dispDW} and comparing it with \cref{eq:DImcc}, one gets
\begin{align}\label{eq:f2pcc}
    \lambda_c \,f^2(p)=\frac{(2\pi)^3h'_p}{-\pi\,p^2}\exp\left(-\frac{1}{\pi}\mathcal{P}\int dE \frac{\delta(E)}{E-h_p}\right)\sin\delta(h_p)\,C(h_p) \,.
\end{align}

One can go further when there is a bound state:
\begin{align}
    \ket{B} = \frac{N}{E_B-H_0} \ket{f}  \,,
\end{align}
with the normalization factor
\begin{align}
    N = \bra{f}\frac{1}{(E_B-H_0)^2}\ket{f}^{-1/2} \,.
\end{align}
This factor also shows up in the derivative of the Fredholm determinant:
\begin{align}
    \frac{d}{d\,W} D(W) \Big{|}_{W=E_B} = \lambda_c \bra{f}\frac{1}{(E_B-H_0)^2}\ket{f} = \exp\left(- \frac{1}{\pi} \int dE \frac{\delta(E)}{E-E_B} \right) \frac{1}{E_B-E_\text{th}} \,.
\end{align}
So, one ends up with
\begin{align}\label{eq:wfcc}
    |\braket{p|B}|^2 &= \frac{N^2}{(E_B-h_p)^2} f^2(p) = \frac{E_B-E_\text{th}}{(E_B-h_p)^2}\exp\left( \frac{1}{\pi} \int dE \frac{\delta(E)}{E-E_B} \right) \lambda_c\,f^2(p) \nonumber\\
    &=\frac{-(E_B-E_\text{th})}{(h_p-E_\text{th})(h_p-E_B)} \frac{(2\pi)^3h'_p}{\pi\,p^2} \sin\delta(h_p)\exp\left[-\frac{1}{\pi}\mathcal{P}\int dE \left(\frac{\delta(E)}{E-h_p}-\frac{\delta(E)}{E-E_B}\right)\right] \,.
\end{align}

\subsection{Model ``bc''}\label{sec:bc}
A general Hamiltonian with a single bare state can have the following form:
\begin{align}\label{eq:rwH}
    H = H_0 + V &= \left[ m \ket{b}\bra{b} + \int \frac{p^2 dp}{(2\pi)^3} h_p \ket{p}\bra{p} \right] \nonumber\\
    &+  \Big[ \lambda_{b} \left( \ket{b}\bra{f} + \ket{f}\bra{b} \right) + \int \frac{p^2 dp}{(2\pi)^3} \frac{k^2 dk}{(2\pi)^3} V(p,k) \ket{p}\bra{k}  \Big] \,.
\end{align}
The model ``bc'' can serve as the bare-state-dominance approximation.
This approximation ignores the $V(p,k)$ term in \cref{eq:rwH}, and leads us to a Hamiltonian of the form of the model ``bc'' \cref{eq:Hbc}.
The Bloch-Horowitz theory \cite{Bloch:1958determination,Luu:2005qr} (which we review in \cref{app:BH}) allows one to integrate out the bare state in \cref{eq:rwH}; then, the bare-continuum interaction can be effectively incorporated into an energy-dependent potential:
\begin{align}\label{eq:Vebc}
    V_{\text{eff}}(p,k;E) = \frac{\lambda_b^2\,f(p)\,f(k)}{E-m} + V(p,k) \,.
\end{align}
So the bare-state-dominance approximation should be good around energies close to the bare mass.

Now we study the model ``bc''.
The Fredholm determinant of this model is 
\begin{align}\label{eq:Dbc}
    D(W)= 1- \frac{\lambda_b^2}{W-m} F(W) \,,
\end{align}
and its imaginary part is as follows:
\begin{align}\label{eq:DImbc}
    \mathrm{Im}\,D(h_p+i\varepsilon) = \frac{\lambda_b^2}{h_p-m}  \pi \bra{f}\delta_D(h_p-H_0)\ket{f} = \frac{ \pi p^2}{(2\pi)^3 h'_p} \frac{\lambda_b^2\,f^2(p)}{h_p-m} \,.
\end{align}
Now the behavior of the phase shift is
\begin{itemize}
    \item $m-E_\text{th}>-\lambda_b^2\,F(E_\text{th})>0$ (no bound states): $\delta \in [-\pi,0]$ and $\delta(E_\text{th})=-\pi$,
    \item $m-E_\text{th}<-\lambda_b^2\,F(E_\text{th})$ (one bound state): $\delta \in [-\pi,0]$ and $\delta(E_\text{th})=0$.
\end{itemize}

The Fredholm determinant has the dispersive representation \cref{eq:dispDW} with
\begin{align}\label{eq:Cbc}
    C(W) = \begin{cases}
        \frac{W-E_\text{th}}{W-m}&N_B=0\vspace{.2cm}\\
        \frac{W-E_B}{W-m}&N_B=1\\
       \end{cases} \,.
\end{align}
Using the trace formula \cref{eq:trf} and noting that $\tr V=0$, the bare mass is
\begin{align}\label{eq:bmbc}
m=\begin{cases}
    E_\text{th}-\frac{1}{\pi}\int dE\,\delta(E) &N_B=0\,,\\
        E_B-\frac{1}{\pi}\int dE\,\delta(E) &N_B=1\,.\\
    \end{cases}
\end{align}
This formula can be understood in the extremely weak-coupling limit $\lambda_b \to 0$, where $\delta(E)=-\pi\,\theta(m-E)$, where $\theta$ is the Heaviside theta function. 
When $m<E_{\text{th}}$, we have $N_B=1$ and $\delta(E)=-\pi\,\theta(m-E)=0$. 
When $m>E_{\text{th}}$, we have $N_B=0$.
Now by taking the imaginary part of \cref{eq:dispDW} and comparing it with \cref{eq:DImbc}, one gets
\begin{align}\label{eq:f2pbc}
    \lambda_b^2\,f^2(p)=\frac{(2\pi)^3h'_p}{\pi p^2}\exp\left(-\frac{1}{\pi}\mathcal{P}\int dE \frac{\delta(E)}{E-h_p}\right)|\sin\delta(h_p)| \times \begin{cases}
        h_p-E_\text{th} &N_B=0\\
        h_p-E_B &N_B=1\\
       \end{cases}\,.
\end{align}

The bare state is distributed in the energy eigenstates:
\begin{align}
    1 = \braket{b|b} = Z + \int \frac{p^2dp}{(2\pi)^3} |\braket{b|p^+}|^2 \,,
\end{align}
where $\ket{p^+}$ are the scattering `in' states, and $Z$ disappears when no bound states are present.
By solving the Lippmann-Schwinger equation
\begin{align}\label{eq:LS}
    \ket{p^+} = \ket{p} + \frac{1}{h_p-H_0+i\varepsilon}V\ket{p^+} \,,
\end{align}
$\ket{p^+}$ is found to be
\begin{align}\label{eq:LSpbc}
    \ket{p^+} = \ket{p} + \frac{c_1}{h_p-m}\ket{b} + \frac{c_2}{h_p-H_0}\ket{f} \,,
\end{align}
where
\begin{align}\label{eq:c1c2bc}
    c_1 = \frac{\lambda_b\,f(p)}{D(h_p+i\varepsilon)} \,, \qquad c_2 = \frac{c_1 - \lambda_b\,f(p)}{\lambda_b\,F(h_p+i\varepsilon)} \,.
\end{align}
Using \cref{eq:esin,eq:DImbc}, one can get the distribution amplitude,
\begin{align}\label{eq:bsdbc}
    \braket{b|p^+} = \frac{e^{i\delta}\sin\delta}{\frac{-\pi p^2}{(2\pi)^3h_p'}\lambda_b\,f(p)} \,,
\end{align}
and the inverse scattering representation of the distribution,
\begin{align}\label{eq:Pbc}
    |\braket{b|p^+}|^2 = \frac{(2\pi)^3 h_p'}{\pi p^2}\exp\left(\frac{1}{\pi}\mathcal{P}\int dE \frac{\delta(E)}{E-h_p}\right)|\sin\delta(h_p)| \times \begin{cases}
        1/(h_p-E_\text{th}) &N_B=0\\
        1/(h_p-E_B) &N_B=1\\
       \end{cases}\,.
\end{align}
We note that \cref{eq:bsdbc} reveals that the bare-state distribution amplitude is proportional to the $T$ matrix ($\propto e^{i\delta}\sin\delta$).
Therefore, a resonance peak is expected in the distribution. 
In fact, this is also true for a general potential, as discussed in Ref.~\cite{Wu:2016ixr}.

The possible bound state in this model is
\begin{align}
    \ket{B} = \sqrt{Z} \left(\ket{b}+\frac{\lambda_b}{E_B-H_0}\ket{f}\right) \,,
\end{align}
where
\begin{align}\label{eq:Zbe}
    Z = |\braket{b|B}|^2 = \frac{1}{1+\bra{f}\frac{\lambda_b^2}{(E_B-H_0)^2}\ket{f}}
\end{align}
represents the bare-state proportion of the bound state.
$Z$ has a concise dispersive representation,
\begin{align}\label{eq:Zbc}
    Z=\exp\left(\frac{1}{\pi}\int dE \frac{\delta(E)}{E-E_B}\right) \,,
\end{align}
which can be derived by noting that
\begin{align}\label{eq:F2bc}
    \bra{f}\frac{\lambda_b^2}{(E_B-H_0)^2}\ket{f} &= \left(-\frac{\partial}{\partial W}\right) \left[\lambda_b^2\,F(W)\right] \Big{|}_{W=E_B} \nonumber\\
    &= \left(-\frac{\partial}{\partial W}\right) \left[W-m-(W-m)D(W)\right]\Big{|}_{W=E_B} \nonumber\\
    &= \left(-\frac{\partial}{\partial W}\right) \left[W-m-(W-E_B)\exp\left(-\frac{1}{\pi}\int dE \frac{\delta(E)}{E-W}\right) \right]\Big{|}_{W=E_B} \nonumber\\
    &= -1 + \exp\left(-\frac{1}{\pi}\int dE \frac{\delta(E)}{E-E_B}\right) \,.
\end{align}
The wave function now becomes
\begin{align}\label{eq:wfbc}
    |\braket{p|B}|^2 = \frac{1}{h_p-E_B} \frac{(2\pi)^3h'_p}{\pi\,p^2} \exp\left[-\frac{1}{\pi}\mathcal{P}\int dE \left(\frac{\delta(E)}{E-h_p}-\frac{\delta(E)}{E-E_B}\right)\right]|\sin\delta(h_p)| \,.
\end{align}

We note that \cref{eq:Zbc,eq:wfbc} coincide with the formulas derived in Ref.~\cite{Li:2021cue}.
However, this should not come as a surprise because the approximations employed there are exact in the model ``bc''.
We also note that this formula will be identical to \cref{eq:wfcc} if one identifies $\delta_{bc}$ with $\delta_{cc}-\pi$.
This is also reasonable because a ``cc'' model can be recognized as a special case of the ``bc'' model when $m\to\pm\infty$ with $-\frac{\lambda_b^2}{m}$ fixed at $\lambda_c$, as reflected by \cref{eq:Vebc}.
This point was also discovered in Ref.~\cite{Weinberg:1962hj}

Finally, we also consider the shallow bound state, \ie, a state with $E_B$ close to $E_{\text{th}}$.
Because of the absence of the continuum self-interaction, the bare state is doubtless crucial to the generation of the bound state.
Hence $E_B$ should be sensitive to the bare mass $m$, making it easy to implement the limit $E_B\to E_{\text{th}}$ by tuning $m$ with $\lambda_b\,f_b(p)$ fixed.
Then the factor
\begin{align}
    \bra{f}\frac{\lambda_b^2}{(E_B-H_0)^2}\ket{f} = \lambda_b^2 \int \frac{q^2dq}{(2\pi)^3} \frac{f^2(q)}{(E_B-h_q)^2}
\end{align}
diverges in the infrared, and one should see $Z\to0$ from \cref{eq:Zbe}.
So naturally a shallow bound state tends to have a small bare-state proportion, even though the bare state is crucial for its generation.

\subsection{Model ``bcc''}

In the model ``bc'', we ignore the whole continuum self-interaction. 
In the model ``bcc'', however, we retain part of it, and thus this model can serve as a correction to the bare-state-dominance approximation.
To be concrete, one can choose a specific energy $h_{\bar{p}}$ around the bare mass, and set $\lambda_c$ as $V(\bar{p},\bar{p})/f^2(\bar{p})$ or, equivalently, $\braket{\bar{p}|V_{\text{bcc}}|\bar{p}}=\braket{\bar{p}|V|\bar{p}}$.
Then, their $\braket{\bar{p}|T|\bar{p}}$ are matched at the leading order of the Born series.

The Fredholm determinant of this model is 
\begin{align}\label{eq:Dbcc}
    D(W)= 1- \left(\frac{\lambda_b^2}{W-m} + \lambda_c \right) \,F(W) = 1- \frac{E_C-W}{E_C-m} \frac{\lambda_b^2}{W-m}F(W) \,,
\end{align}
where
\begin{align}\label{eq:EC}
    E_C:=m-\frac{\lambda_b^2}{\lambda_c}
\end{align}
is known as the Castillejo-Dalitz-Dyson (CDD) zero \cite{Castillejo:1955ed}, because it corresponds to a zero of the on-shell $T$-matrix element $\braket{p|T(h_p+i\varepsilon)|p}$.
In this special model, it is even a zero of the off-shell $T$-matrix element:
\begin{align}
    \braket{p|T(W)|k}=\frac{\left(\frac{\lambda_b^2}{W-m} + \lambda_c \right)f(p)\,f(k)}{D(W)}=\frac{\frac{E_C-W}{E_C-m} \frac{\lambda_b^2}{W-m}f(p)\,f(k)}{D(W)}\,.
\end{align}
At $E_C$, the system feels no interactions because the bare-continuum interaction cancels the continuum self-interaction.
So the presence of a CDD zero may indicate an interplay between the two interactions.
In the context of hadron physics, bare states typically represent compact multiquark states while continuum states are molecular-like multihadron states.
The presence of a CDD zero is recognized as an interplay of quark and hadron degrees of freedom \cite{Baru:2010ww,Hanhart:2011jz,Guo:2016wpy,Kamiya:2016oao,Kang:2016jxw}.

To analyze the bound states of this model, we introduce 
\begin{align}
    A(W):=\frac{\lambda_b^2}{E_C-m}\frac{E_c-W}{W-m} \,,
\end{align}
so that $D(W)=1-A(W)F(W)$.
The analysis of bound states only cares about real-valued $E<E_{\text{th}}$ where we have $F(E)<0$ and $F'(E)<0$, and hence $F(E)$ is both negative and monotonically decreasing.
If $A(E)>0$, then $D(E)\geq 1$ and no bound states can appear because a bound state's energy satisfies $D(E_B)=0$.
If $A(E)<0$, then by noting that
\begin{align}
    A'(E)=-\frac{\lambda_b^2}{(W-m)^2}\leq 0 \,,
\end{align}
we have
\begin{align}
    D'(E)= -A'(E)F(E)-A(E)F'(E) \leq 0 \,,
\end{align}
which means that $D(E)$ is monotonically decreasing except at $E=m$.
The remaining analysis should be split into six cases depending on the ordering of $E_C$, $m$, and $E_\text{th}$.
The results are listed in \cref{tab:bccBound}, where we use $(a,b,c)$ to denote the case $-\infty<a<b<c<+\infty$.
Taking the first case $(E_C,m,E_{\text{th}})$ as an example, it is easy to find that $D(E<E_C)\geq 1$, $D(E_C)=1$, $D(m_-)=-\infty$, $D(m_+)=+\infty$ and $D(E_\text{th})\geq 1$, so only a single bound state appears and lies in $[E_C,m]$.

From the table, it is easy to summarize that each of the conditions $m<E_\text{th}$ and $D(E_\text{th})<0$ can produce a bound state.
The condition $m<E_\text{th}$ indicates that a below-threshold bare state will evolve into a physical bound state directly.
The condition $D(E_\text{th})<0$ can be equivalently expressed as $\lambda_c-\frac{\lambda_b^2}{m-E_\text{th}}<F(E_\text{th})^{-1}<0$, \ie, a negative $\lambda_c$ and $-\frac{\lambda_b^2}{m-E_\text{th}}$ tend to form a bound state.
A negative $\lambda_c$ corresponds to an attractive continuum self-interaction, and a negative $-\frac{\lambda_b^2}{m-E_\text{th}}$ corresponds to the bare-continuum interaction with an above-threshold bare state.
So there are three different ways of forming a bound state: directly by a below-threshold bare state, by an attractive continuum self-interaction, and by the bare-continuum interaction with an above-threshold bare state.
On the other hand, we can rewrite \cref{eq:EC} as $\lambda_c=\frac{\lambda_b^2}{m-E_C}$, using which we can know if the continuum self-interaction is attractive or repulsive by looking at the sign of $m-E_C$.

Now we look at the three cases satisfying $m<E_\text{th}$, where at least a bound state is present, and the bare-continuum interaction resists the generation of the other bound state. 
The cases $(E_C,m,E_\text{th})$ and $(m,E_\text{th},E_C)$ have repulsive and weak-attractive continuum self-interactions respectively. 
Neither can form the second bound state.
The case $(m,E_C,E_\text{th})$ has a strong-attractive continuum self-interaction. 
Two bound states appear only if it is strong enough. This is the only case where two bound states can be formed.
If we now turn off the bare-continuum interaction, \ie, we set $\lambda_b=0$, we still have two bound states. 
One is simply the below-threshold bare state, and the other is a molecular-like state purely generated by the continuum self-interaction.
After we turn on the bare-continuum interaction, the bare state and the molecular-like state are mixed to form two new bound states.

Then, we look at the three cases satisfying $m>E_\text{th}$. Now the bare-continuum interaction behaves as an attractive interaction in forming a bound state.
The case $(E_\text{th},E_C,m)$ has a strong-repulsive continuum self-interaction, and no bound states can be formed.
The case $(E_C,E_\text{th},m)$ has a weak-repulsive continuum self-interaction. 
If the bare-continuum interaction is strong enough, a bound state can be formed.
The case $(E_\text{th},m,E_C)$ has an attractive continuum self-interaction. 
A bound state can be formed only if the overall attraction of the two interactions is strong enough.
Which interaction dominates the generation of the bound state depends on the relative strength of the two interactions.
The ratio $\lambda_c\Big{/}\left(-\frac{\lambda_b^2}{m-E_\text{th}}\right)=\frac{m-E_\text{th}}{E_C-m}$ directly provides a way to quantify the generation mechanism of the bound state. 
However, there are several flaws. 
First, the ratio does not have a direct physical meaning. 
Second, the definition of the ratio relies on the model ``bcc'', and cannot be generalized. 
Last, the ratio does not depend on the bound state itself, \ie, it reflects only the relative strength between the two interactions, instead of their impact on the bound state.

With the analysis of bound states, the Fredholm determinant will have the dispersive representation \cref{eq:dispDW} with
\begin{align}\label{eq:Cbcc}
    C(W) = \begin{cases}
        \frac{W-E_\text{th}}{W-m}&N_B=0\vspace{.2cm}\\ 
        \frac{W-E_B}{W-m}&N_B=1\vspace{.2cm}\\ 
        \frac{(W-E_{B_1})(W-E_{B_2})}{(W-E_\text{th})(W-m)}&N_B=2\\
       \end{cases} \,.
\end{align}

\begin{table}[tbp]
    \centering
    \caption{Bound states of model ``bcc'' in different cases, where  $(a,b,c)$ represents $-\infty<a<b<c<+\infty$.}\label{tab:bccBound}
    \renewcommand\arraystretch{1.5}
    \begin{ruledtabular}
    \begin{tabular}{llc}
        Case & Bound states \\ \hline
        $(E_C,m,E_\text{th})$ & One in $[E_C,m]$. \\
        $(m,E_C,E_\text{th})$ & One in $(-\infty,m]$. One in $[E_C,E_\text{th}]$ if $D(E_\text{th})<0$. \\
        $(E_C,E_\text{th},m)$ & One in $[E_C,E_\text{th}]$ if $D(E_\text{th})<0$. \\
        $(m,E_\text{th},E_C)$ & One in $(-\infty,m]$.\\
        $(E_\text{th},E_C,m)$ & No bound states. \\
        $(E_\text{th},m,E_C)$ & One in $(-\infty,E_\text{th}]$ if $D(E_\text{th})<0$. \\
    \end{tabular}
    \end{ruledtabular}
\end{table}

When $E_C$ is above the threshold, the phase shift at $E_C$ is a multiple of $\pi$.
Recalling the definition \cref{eq:ddelta} and $D(E_C)=D(\infty)=1$, we have
\begin{align}
    \delta(E_C) = \arg \frac{C(E_C+i\varepsilon)}{C(\infty+i\varepsilon)} = - \arg \frac{E_C+i\varepsilon-m}{\infty +i \varepsilon -m} =\begin{cases}
    0 &\text{$E_C>m$}\\
    -\pi &\text{$E_C<m$}\\
    \end{cases}  \,,
\end{align}
where the convention $\delta(\infty)=0$ is used.
So only the case $(E_\text{th},E_C,m)$, which has no bound states, can have $\delta(E_C)=-\pi\neq0$.
When $E_C$ is below the threshold, one has to locate it as a zero of the $T$ matrix using analytic continuation.
For other behaviors of the phase shift, the discussion is much more complicated than in previous models, so we will not analyze them here.

Now because $\tr V = \lambda_c \braket{f|f}\neq 0$, the trace formula \cref{eq:trf} does not provide a useful expression for the bare mass as in \cref{eq:bmbc}.
Instead, one should look at 
\begin{align}\label{eq:DEC}
    D(E_C\pm i\varepsilon)=1=s\exp\left(-\frac{1}{\pi}\int dE \frac{\delta(E)}{E-E_C}\right)C(E_C) \,,
\end{align}
where
\begin{align}
    s=\begin{cases}
    -1 &\text{$E_\text{th}<E_C<m$}\\
    +1 &\text{Others}\\
    \end{cases} \,.
\end{align}
Then we get the bare mass
\begin{align}\label{eq:bmbcc}
    m = E_C - s\exp\left(-\frac{1}{\pi}\int dE \frac{\delta(E)}{E-E_C}\right) C_1(E_C) \,,
\end{align}
where $C_1$ is defined as
\begin{align}
    C_1(W) := (W-m)C(W) = \begin{cases}
        W-E_\text{th}&N_B=0\\ 
        W-E_B&N_B=1\\ 
        (W-E_{B_1})(W-E_{B_2})/(W-E_\text{th})&N_B=2\\
       \end{cases} \,.
\end{align}

Now we come back to the inverse problem. We first work out the imaginary part of $D$:
\begin{align}\label{eq:DImbcc}
    \text{Im}\,D(h_p+i\varepsilon) = \frac{\pi p^2}{(2\pi)^3 h'_p} \frac{E_C-h_p}{E_C-m} \frac{\lambda_b^2\,f^2(p)}{h_p-m}  \,.
\end{align}
Then, comparing it with the imaginary part of the dispersive representation, we get
\begin{align}\label{eq:f2pbcc}
    \lambda_b^2\,f^2(p) &= \frac{(2\pi)^3h'_p}{\pi p^2} \frac{E_C-m}{E_C-h_p} \exp\left(-\frac{1}{\pi}\mathcal{P}\int dE \frac{\delta(E)}{E-h_p}\right)[-\sin\delta(h_p)] C_1(h_p) \nonumber\\
    &=  \frac{(2\pi)^3h'_p}{\pi p^2} \frac{C_1(E_C)}{E_C-h_p} s \exp\left[-\frac{1}{\pi}\mathcal{P}\int dE \left(\frac{\delta(E)}{E-h_p}-\frac{\delta(E)}{E-E_C}\right)\right][-\sin\delta(h_p)] C_1(h_p)  \,,
\end{align}
where in the second line we have used \cref{eq:bmbcc}.

The bare state is distributed in the energy eigenstates:
\begin{align}
    1 = \braket{b|b} = \sum_i Z_i + \int \frac{p^2dp}{(2\pi)^3} |\braket{b|p^+}|^2 \,,
\end{align}
where $\ket{p^+}$ is the scattering `in' state, and $Z_i:=|\braket{b|B_i}|^2$ is the bare-state proportion of the bound state $\ket{B_i}$.
By solving the Lippmann-Schwinger equation \cref{eq:LS}, $\ket{p^+}$ can be found to be
\begin{align}\label{eq:LSpbcc}
    \ket{p^+} = \ket{p} + \frac{c_1}{h_p-m}\ket{b} + \frac{c_2}{h_p-H_0}\ket{f} \,,
\end{align}
where
\begin{align}\label{eq:c1c2bcc}
    c_1 = \frac{\lambda_b\,f(p)}{D(h_p+i\varepsilon)} \,, \qquad c_2 = \frac{c_1 - \lambda_b\,f(p)}{\lambda_b\,F(h_p+i\varepsilon)} \,.
\end{align}
Note that, although \cref{eq:LSpbcc,eq:c1c2bcc} have exactly the same form as \cref{eq:LSpbc,eq:c1c2bc}, $D$ and $F$ are different functions.
Using \cref{eq:esin,eq:DImbcc}, one can first get the distribution amplitude,
\begin{align}
    \braket{b|p^+} = \frac{e^{i\delta}\sin\delta}{\frac{-\pi p^2}{(2\pi)^3h_p'} \frac{E_C-h_p}{E_C-m} \lambda_b\,f(p)} \,,
\end{align}
and then the inverse scattering representation of the distribution:
\begin{align}\label{eq:Pbcc}
    |\braket{b|p^+}|^2 &= \frac{(2\pi)^3h_p'}{\pi p^2}\frac{E_C-m}{E_C-h_p} \exp\left[\frac{1}{\pi}\mathcal{P}\int dE \left(\frac{\delta(E)}{E-h_p}\right)\right] [-\sin\delta(h_p)]C_1(h_p) \nonumber\\
    &= \frac{(2\pi)^3h_p'}{\pi p^2}\frac{C_1(E_C)}{E_C-h_p} s \exp\left[\frac{1}{\pi}\mathcal{P}\int dE \left(\frac{\delta(E)}{E-h_p}-\frac{\delta(E)}{E-E_C}\right)\right] [-\sin\delta(h_p)]C_1(h_p)\,.
\end{align}

From now on, we focus only on the single bound-state case.
The bound state can be solved to be
\begin{align}\label{eq:Bbc}
    \ket{B} = \sqrt{Z}\left( \ket{b} + \frac{\lambda_{B}}{E_B-H_0}\ket{f} \right) \,,
\end{align}
where $\lambda_{B}:=\lambda_b + \frac{\lambda_c}{\lambda_b}(E_B-m)$ and
\begin{align}
    Z = |\braket{b|B}|^2 = \frac{1}{1+\braket{f|\frac{\lambda_B^2}{(E_B-H_0)^2}|f}} \,.
\end{align}
By noting that
\begin{align}
    \bra{f}\frac{\lambda_B^2}{(E_B-H_0)^2}\ket{f} &= \left(-\frac{\partial}{\partial W}\right) \left[\lambda_B^2\,F(W)\right] \Big{|}_{W=E_B} \nonumber\\
    &= \lambda_B^2 \left(-\frac{\partial}{\partial W}\right) \left[\frac{W-m-(W-m)D(W)}{\lambda_b^2+\lambda_c(W-m)} \right]\Big{|}_{W=E_B} \nonumber\\
    &= -1 + \frac{\lambda_B}{\lambda_b} \exp\left(-\frac{1}{\pi}\int dE \frac{\delta(E)}{E-E_B}\right) \,,
\end{align}
and
\begin{align}
    \frac{\lambda_B}{\lambda_b} & = \frac{E_C-E_B}{E_C-m} = \exp\left(\frac{1}{\pi}\int dE \frac{\delta(E)}{E-E_C}\right) \,,
\end{align}
one finds the dispersive representation of $Z$:
\begin{align}\label{eq:Zbcc}
    Z = \exp\left[\frac{1}{\pi}\int dE \left(\frac{\delta(E)}{E-E_B}-\frac{\delta(E)}{E-E_C}\right)\right] \,.
\end{align}
It is now straightforward to get the wave function:
\begin{align}\label{eq:wfbcc}
    |\braket{p|B}|^2 = \frac{1}{h_p-E_B} \frac{(2\pi)^3h'_p}{\pi\,p^2} &\frac{(E_C-E_B)^3}{(E_C-m)^2(E_C-h_p)}\nonumber\\
    &\times \exp\left[-\frac{1}{\pi}\mathcal{P}\int dE \left(\frac{\delta(E)}{E-h_p}-\frac{\delta(E)}{E-E_B}\right)\right] [-\sin\delta(h_p)] \,.
\end{align}

At the end of \cref{sec:bc}, we showed that the bare-state proportion $Z$ cannot faithfully describe the generation of the bound state.
In the ``bc'' model, while the bound state is doubtless generated by the bare state, it can still happen that $Z\sim 0$.
Now, in the ``bcc'' model, both the bare-continuum interaction and continuum self-interaction are included, so it is natural to ask which interaction dominates the generation of the bound state.
Using \cref{eq:EC}, we know that $E_C<m$ ($E_C>m$) indicates a repulsive (attractive) continuum self-interaction, and a larger separation between them indicates a weaker continuum self-interaction or, equivalently, a stronger bare-continuum interaction.
However, this not only does not provide a quantitative criterion, but also has nothing to do with the bound state.

A way to quantify this is to consider the decomposition of the Hamiltonian \cref{eq:Hbcc}.
The Hamiltonian can be split into four parts,
\begin{align}
    H = H_0 + V = H_{0b} + H_{0c} + V_{bc} + V_{cc} \,,
\end{align}
and similarly for the bound-state energy $E_B=\braket{B|H|B}$, \ie, $\braket{B|H_{0b} + H_{0c} + V_{bc} + V_{cc}|B}\equiv E_{0b}+E_{0c}+E_{bc}+E_{cc}$.
To quantify the interplay between the two interactions inside the bound state, we can introduce the potential ratio:
\begin{align}
    R:=\frac{E_{cc}}{E_{bc}} \,.
\end{align}

The first three terms of the energy decomposition can be model-independently expressed as follows:
\begin{align}\label{eq:mi3}
    E_{0b} &= Z m \,, \nonumber\\
    E_{0c} &= \braket{B|H_0 P_X|B} = X E_B - \braket{B|V P_X|B} \,, \nonumber\\
    &= X E_B - E_{bc}/2 - E_{cc} \,, \nonumber\\
    E_{bc} &= E_B - (E_{0b}+E_{0c}+E_{cc}) = E_B - Zm -XE_B +E_{bc}/2 \nonumber\\
    &= -2Z(m-E_B) \,,
\end{align}
where $P_X$ is the projection operator for continuum states, and $X=1-Z$.

When the bare mass is located above the threshold, the bound-state energy is of course smaller than the bare mass. 
When the bare mass is located below the threshold, according to \cref{tab:bccBound}, there must be a bound state present below the bare mass.
So for the single-bound-state case, the bare mass has to be located above the bound-state energy. 
Then we always have $E_{bc}<0$, which means that the bare-continuum interaction always has an attractive effect in forming the bound state.
Then, by noting that
\begin{align}
    E_{0c} &=  \braket{B|H_0 P_X|B} \geq X E_{\text{th}} \,, \nonumber\\
    E_{cc} &\leq E_B - (E_{0b} + X E_{\text{th}} + E_{bc}) = -X(E_{\text{th}}-E_B) + Z(m-E_B) \leq Z(m - E_B) \,,
\end{align}
we can derive an inequality for the potential ratio:
\begin{align}\label{eq:Rie}
    R = \frac{E_{cc}}{E_{bc}} \geq \frac{Z(m-E_B)}{-2Z(m-E_B)} = -\frac{1}{2} \,,
\end{align}
or $E_{cc} \leq -E_{bc}/2$.
This inequality puts a general upper limit on the repulsive strength of the continuum self-interaction, exceeding which a bound state is unable to be formed.
We emphasize that \cref{eq:Rie} holds model independently whenever the bound state appears below a bare state.

The last term of the energy decomposition should be estimated using the solution \cref{eq:Bbc}:
\begin{align}
    E_{cc} = Z \lambda_c \lambda_B^2 F^2(E_B) = -Z \frac{(m-E_B)^2}{E_C-m} \,,
\end{align}
where we have used
\begin{align}
    D(E_B)=0 \;\Rightarrow\; F(E_B)=\frac{E_B-m}{\lambda_B\lambda_b} \,.
\end{align}
Then the potential ratio becomes
\begin{align}\label{eq:prR}
    R=\frac{E_{cc}}{E_{bc}}=\frac{1}{2}\frac{m-E_B}{E_C-m} \,.
\end{align}
So by comparing twice the distance between $E_C$ and $m$ with the distance between $E_B$ and $m$, one can determine which dominates the generation of the bound state.
In hadron physics, there are many states, like $D_{s0}^*(2317)$ and $X(3872)$, that have masses far away from the corresponding quark-model predictions. 
For such a state, its observed mass is simply $E_B$, and we can treat the quark-model prediction as the bare mass $m$. Then, locating the CDD zero can tell us the generation mechanism of the state.
In addition, the potential ratio can also be defined for scattering states:
\begin{align}
    R(h_p):=\frac{\braket{p^+|V_{cc}|p^+}}{\braket{p^+|V_{bc}|p^+}}= \frac{1}{2}\frac{m-h_p}{E_C-m} \,.
\end{align}

\section{Analysis of real-world systems}\label{sec:rws}

\subsection{Deuteron in proton-neutron system}\label{sec:deu}
The deuteron is a shallow bound state of the proton-neutron system.
Its properties have been widely studied in hundreds of works (see Ref.~\cite{Zhaba:2017syr} for a review).
After partial-wave projection, the relevant channels are $^3S_1$ and $^3D_1$.
As the $^3D_1$ channel is only expected to have around a 5\% contribution, it is reasonable to consider only the $^3S_1$ channel in the first approximation.

The low-energy $^3S_1$ phase shift can be well described by the effective range expansion \cite{Klarsfeld:1984es}
\begin{align}
    p \cot \delta_\text{ERE} = \frac{1}{a} + \frac{r}{2}p^2
\end{align}
with 
\begin{align}
    a = -5.419(7)\,\text{fm} \,, \quad r = 1.766(8)\,\text{fm} \,.
\end{align}
As $\delta_\text{ERE}(0)- \delta_\text{ERE}(\infty) = \pi$, the Levinson theorem requires the absence of bare states.
Of course, the theorem relies on the high-energy behavior of the phase shift, which is inaccessible in reality.
However, it is still reasonable to expect that the possible bare state does not play a significant role, even though we cannot completely rule out the possibility of its existence.
Therefore, we consider using the model ``cc'' to analyze the deuteron.

As discussed in \cref{app:assep}, the model ``cc'' can approximate any $S$-wave near-threshold physics with an error of $\mathcal{O}\left( \beta^3/M_V^3  \right)$. 
For the deuteron case, we can take $\beta$ to be its binding momentum $b=46\,$MeV, and we also take $M_V=m_\pi=138\,$MeV. 
Then the error is $\mathcal{O}\left( 4\% \right)$.

On the other hand, Weinberg has developed a method \cite{Weinberg:1965zz} for quantitatively estimating the possible bare-state proportion of the deuteron.
The method predicts an unphysical value $Z=-68\%$ with an uncertainty of $\mathcal{O}(b/m_\pi=33\%)$.
Recently, the method was improved in Ref.~\cite{Li:2021cue}. 
It predicts $Z=0\%$ with an uncertainty of $\mathcal{O}(b^2/m_\pi^2=11\%)$.
The authors of Ref.~\cite{Li:2021cue} also derived an expression for the deuteron wave function, which happens to be exactly the same as \cref{eq:wfcc} if one identifies $\delta_{B}$ in that paper with $\delta-\pi$ in \cref{eq:wfcc}.
Therefore, for the deuteron wave function, the model ``cc'' predicts exactly the same result as that presented in Ref.~\cite{Li:2021cue}, except that the uncertainty changed from $\mathcal{O}(b^2/m_\pi^2=11\%)$ to $\mathcal{O}(b^{3}/m_\pi^{3}=4\%)$.
However, we emphasize that such a change in the uncertainty is not an improvement, because the absence of bare states is assumed {\it a priori} in the model ``cc''.

\subsection{$\rho$ meson in $\pi\pi$-$(q\bar{q})$ system}\label{sec:ppr}
The observed phase shift of the $\rho$ meson can be well described by a Breit-Wigner (BW) fit up to at least around $1.2$\,GeV \cite{Werner:2019hxc,Protopopescu:1973sh} (although the $K\bar{K}$ channel appears already below $1.2$\,GeV, the resulting inelasticity is very small, and thus we do not consider the $K\bar{K}$ channel here):
\begin{align}
    \delta_{\text{BW}}(E) = \arctan\left[ \frac{g^2}{6\pi} \frac{p^3}{E\,(m_\rho^2-E^2)} \right] \,,\qquad p = \sqrt{E^2/4-m_\pi^2} \,,
\end{align}
where we take the Particle Data Group (PDG) values \cite{ParticleDataGroup:2020ssz}: $m_\pi=140\,$MeV, $m_\rho=775\,$MeV and $g=5.98$. The coupling $g$ is related to the width $\Gamma_\rho=149\,$MeV via
\begin{align}
    \Gamma_\rho = \frac{g^2}{6\pi} \frac{p^3_\rho}{m_\rho} \,, \qquad p_\rho=\sqrt{m_\rho^2/4-m_\pi^2} \,.
\end{align}
This phase shift satisfies $\delta_{\text{BW}}(0)- \delta_{\text{BW}}(\infty) = -\pi$.
As there are no bound states, the Levinson theorem requires the existence of a single bare state.
Therefore, it excludes the applicability of our ``cc'' model.

Now we consider the ``bc'' model. 
Substituting $\delta_{\text{BW}}$ into the bare mass formula \cref{eq:bmbc} simply gives $m=\infty$, because $\delta_{\text{BW}}(\infty) \neq 0$.
However, this result should not be accepted as true, because the high-energy behavior of the phase shift is crucial in deriving it.
As mentioned before, the BW fit works at least up to $1.2\,$GeV.
Thus, what \cref{eq:bmbc} really tells us is that
\begin{align}
    m \geq E_\text{th} -\frac{1}{\pi}\int_{E_{\text{th}}}^{1.2\,\text{GeV}} dE\,\,\delta_{\text{BW}}(E) = 812\,\text{MeV} \,,
\end{align}
where the inequality holds because $\delta\leq 0$ in the model ``bc''.
On the other hand, the model ``bc'' can be understood as a result of the bare-state-dominance approximation.
We do not expect such an approximation to work in the high-energy region.
So replacing the BW fit with the real-world phase shift up to a higher energy is not a solution for locating the bare mass.
Instead, we choose to model the high-energy behavior of the phase shift with the following replacement
\begin{align}\label{eq:g2bc}
    g^2\to g^2(p)=g^2\left(\frac{1+\frac{p_\rho^2}{\Lambda^2}}{1+\frac{p^2}{\Lambda^2}}\right)^2 \,,
\end{align}
where $g^2(p_\rho)=g^2$ ensures that the near $\rho$-mass behavior of the phase shift does not change too much.
In fact, the main difference between phase shifts with different $\Lambda$ is the high-energy behavior.
We use three different $\Lambda$, and list the corresponding bare masses in \cref{tab:lma}.
From the table, the bare mass is expected to have an uncertainty of $\mathcal{O}(100\,\text{MeV})$ coming from the high-energy behavior of the phase shift.

\begin{table}[tbp]
    \centering
    \caption{The bare mass in different cases.}\label{tab:lma}
    \renewcommand\arraystretch{1.5}
    \begin{ruledtabular}
    \begin{tabular}{cccc}
        $\Lambda$ (MeV) & 800 & 1200 & 1600 \\ \hline
        $m$ (MeV) with $E_C=\infty$ (Model ``bc'') & 846 & 895 & 943 \\
        $m$ (MeV) with $E_C=m_\rho-1000\,$MeV & 846 & 898 & 950 \\
        $m$ (MeV) with $E_C=m_\rho+1000\,$MeV & 846 & 891 & 932 \\
    \end{tabular}
    \end{ruledtabular}
\end{table}

\begin{figure}[tbp]
    \centering
    \includegraphics[width=\columnwidth*24/50]{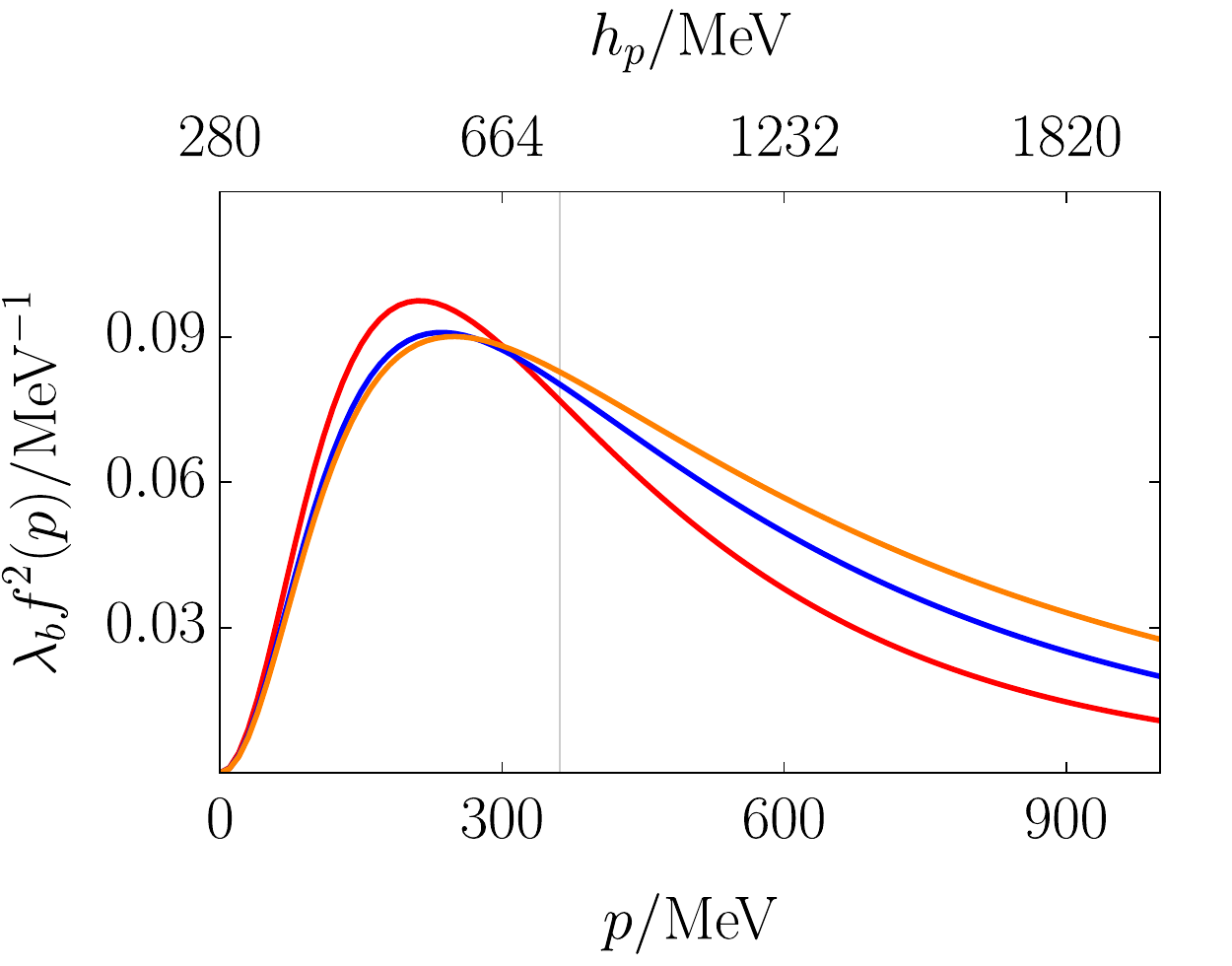}
    \includegraphics[width=\columnwidth*24/50]{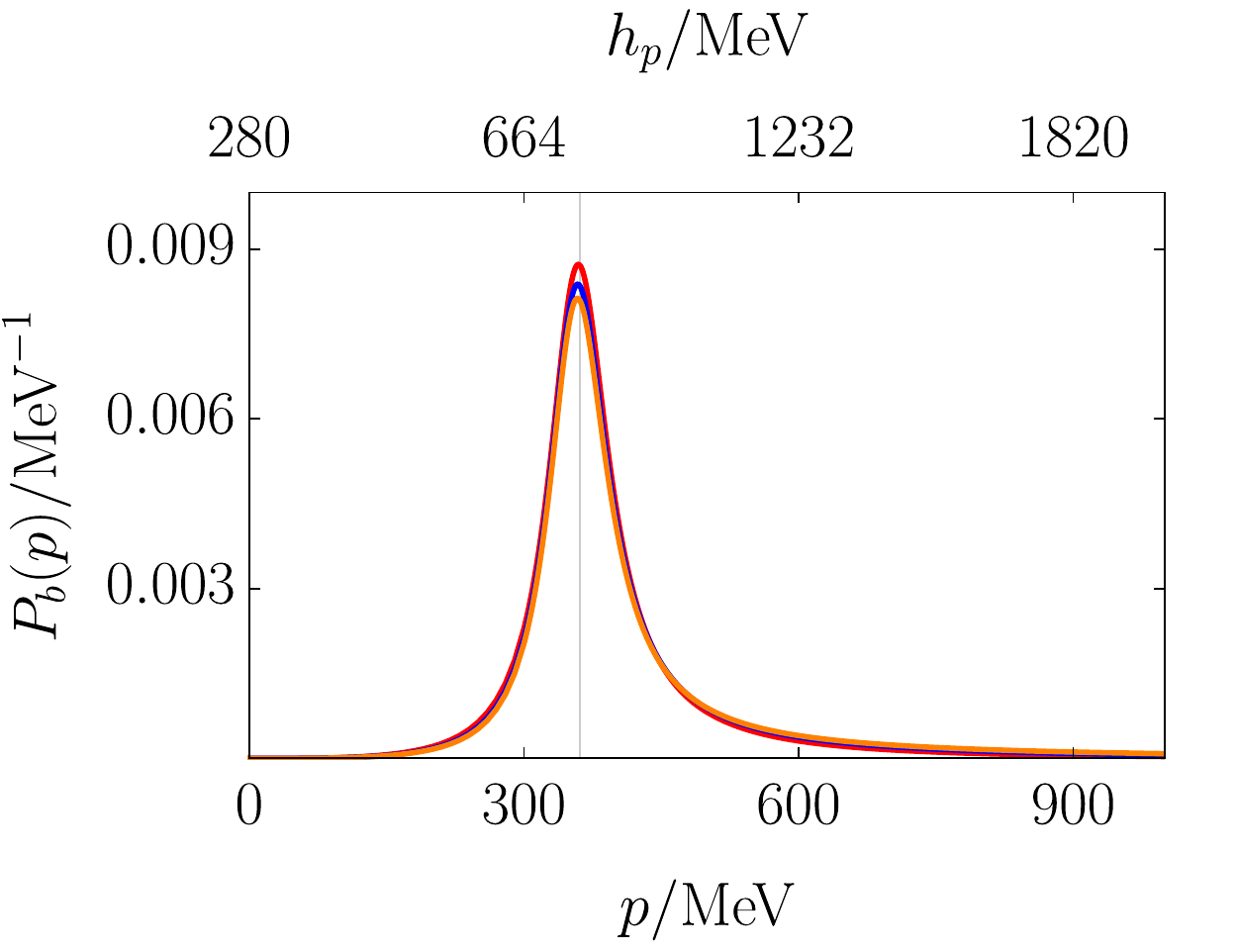}
    \caption{Plots of $\lambda_b^2\,f^2(p)$ and $P_b(p):=\frac{p^2}{(2\pi)^3}|\braket{b|p^+}|^2$ with $\Lambda=$ 800(red), 1200(blue), 1600(orange) MeV. The grey vertical line denotes $h_p=m_\rho$.}\label{fig:fPbc}
\end{figure}

By using \cref{eq:f2pbc,eq:Pbc} with these different values of $\Lambda$, we calculate the corresponding bare-continuum coupling form factor $\lambda_b^2\,f^2(p)$ and
the bare-state distribution $\frac{p^2}{(2\pi)^3}|\braket{b|p^+}|^2$
as shown in \cref{fig:fPbc}.
While variation of $\Lambda$ affects $\lambda_b^2\,f^2(p)$ a bit, the bare state distribution is quite stable.

Next, we consider the model ``bcc''.
The model requires the existence of a CDD zero.
To see how a possible CDD zero may affect our previous predictions, we introduce both $\Lambda$ and a CDD zero into the BW fit with the following replacement
\begin{align}\label{eq:g2bcc}
    g^2\to g^2(p)=g^2\left(\frac{1+\frac{p_\rho^2}{\Lambda^2}}{1+\frac{p^2}{\Lambda^2}}\right)^2\frac{h_p-E_C}{m_\rho-E_C} \,.
\end{align}
The observed phase shift does not suggest that such a zero is present close to the range from $E_\text{th}$ to 1200\,MeV.
We tentatively use $E_C=m_\rho\pm1000\,$MeV to study the effects of $E_C$.
Using \cref{eq:bmbcc}, we include the corresponding bare mass in \cref{tab:lma}.
From the table, the possible CDD zero is only expected to contribute an uncertainty of $\mathcal{O}(10\,\text{MeV})$ to the bare mass.
This uncertainty is much smaller than that under variation of $\Lambda$.

\begin{figure}[tbp]
    \centering
    \includegraphics[width=\columnwidth*24/50]{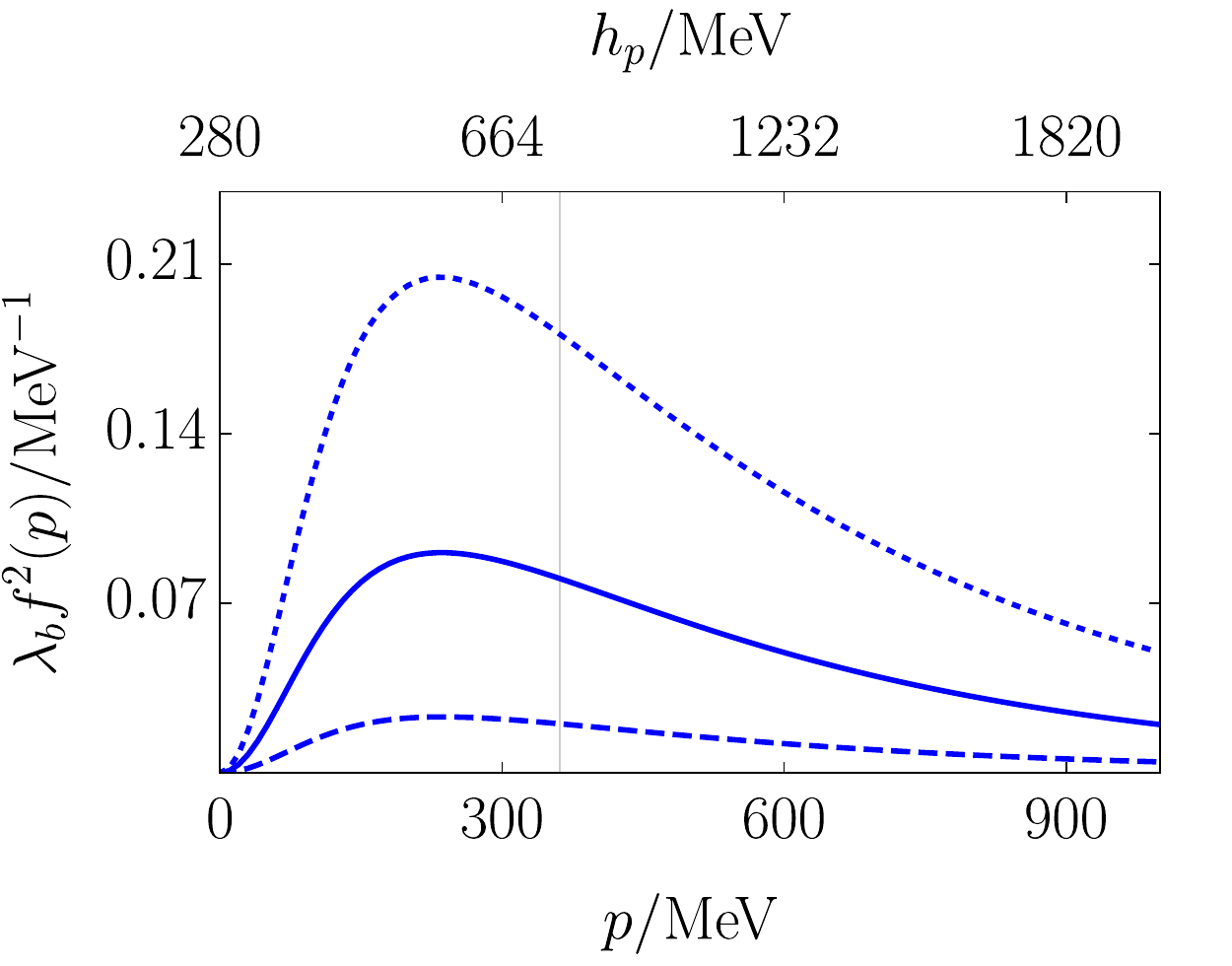}
    \includegraphics[width=\columnwidth*24/50]{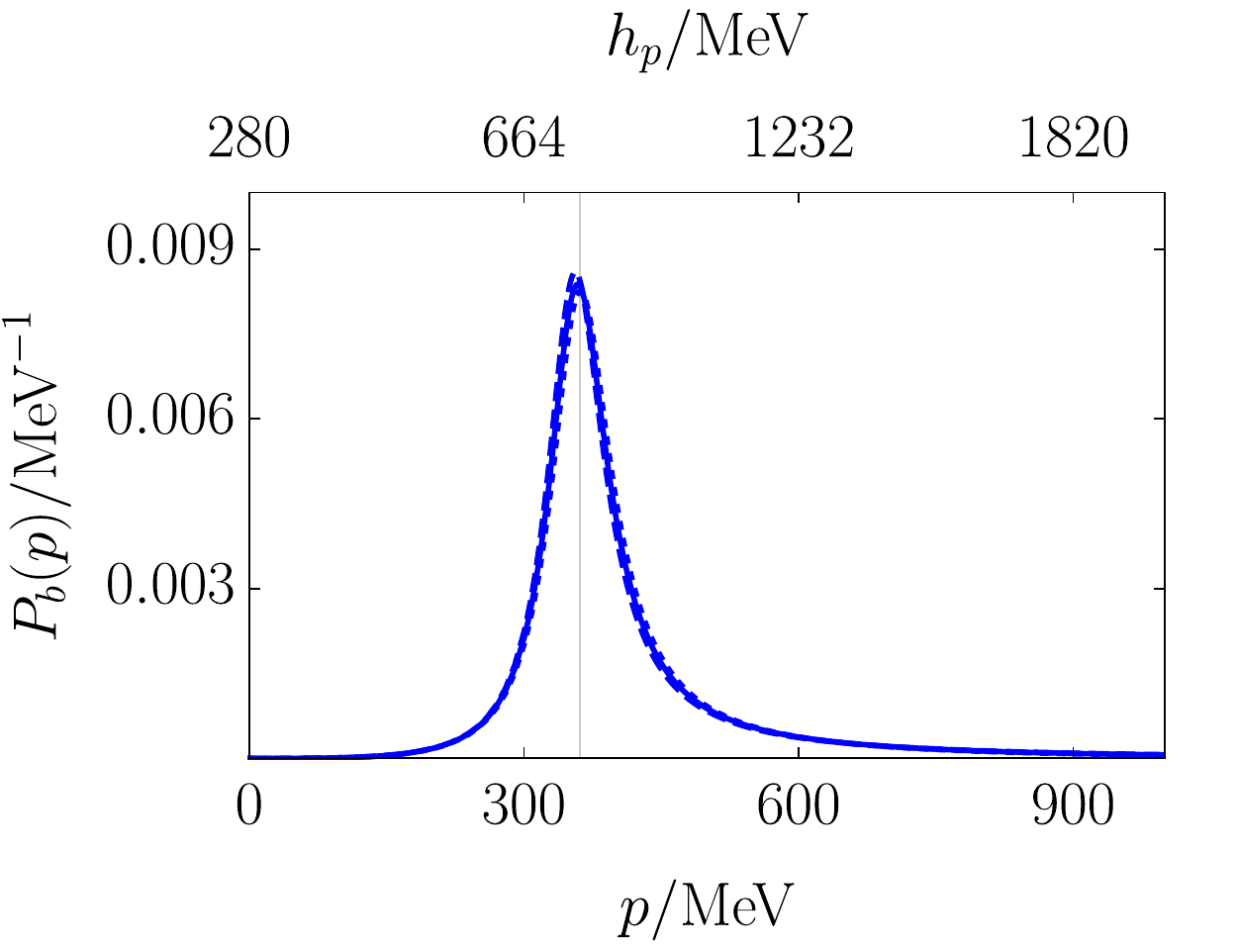}
    \caption{Plots of $\lambda_b^2\,f^2(p)$ and $P_b(p):=\frac{p^2}{(2\pi)^3}|\braket{b|p^+}|^2$ with $\Lambda=1200\,$MeV and $E_C=$ $m_\rho-1000\,$MeV(dashed), $\infty$(solid), $m_\rho+1000\,$MeV(dotted). The grey vertical line denotes $h_p=m_\rho$.}\label{fig:fPbcc}
\end{figure}

Then, we focus on the $\Lambda=1200\,$MeV case.
In \cref{fig:fPbcc} we plot the corresponding $\lambda_b^2\,f^2(p)$ and $\frac{p^2}{(2\pi)^3}|\braket{b|p^+}|^2$ using \cref{eq:f2pbcc,eq:Pbcc}.
We see that while $E_C$ has a drastic effect on $\lambda_b^2\,f^2(p)$, the bare-state distribution is still very stable.

From the above discussions, we have some simple conclusions.
First, the bare mass is expected to be larger than $812\,$MeV, and its precise location is more sensitive to $\Lambda$ than $E_C$.
The former is related to the bare-continuum interaction, while the latter is related to the interplay between the bare-continuum interaction and continuum self-interaction.
Second, the bare-continuum interaction is sensitive to both $\Lambda$ and $E_C$. Thus it can hardly be determined model independently.
Third, the bare-state distribution is quite stable under variations of both $\Lambda$ and $E_C$.
Therefore its extractions in \cref{fig:fPbc,fig:fPbcc} are reliable.

\subsection{$D_{s0}^*(2317)$ in $DK$-$(c\bar{s})$ system}\label{sec:ds0}
$D_{s0}^*(2317)$ can be viewed as a shallow bound state of the $DK$-$(c\bar{s})$ system.
There we have the PDG values \cite{ParticleDataGroup:2020ssz}
\begin{align}
    E_{\text{th}} = 2363\,\text{MeV}\,,\; E_B=2318\,\text{MeV}\,.
\end{align}
For the bare $(c\bar{s})$ mass, the well-known Godfrey-Isgur model \cite{Godfrey:1985xj} predicts $m = 2480 \,\text{MeV}$. 
On the other hand, a recently published work \cite{Yang:2021tvc} followed the Godfrey-Isgur model with updated data, and intentionally removed those masses of near-threshold states as input. 
It predicted $m = 2406 \,\text{MeV}$.

The $Z$ factor of $D_{s0}^*(2317)$ has been investigated in many works \cite{Torres:2014vna,Bali:2017pdv,Cheung:2020mql,Matuschek:2020gqe} based on Weinberg's relations, and tends to have a small central value but with a large systematic uncertainty.
However, whatever the value of $Z$ is, the potential ratio which measures how $D_{s0}^*(2317)$ is generated is still worth studying.
In the following, we will focus on this ratio.

Before using the ``bcc'' model, we first study the system model independently.
Because the bare mass lies above the bound-state energy, we have $R\geq -\frac{1}{2}$ from \cref{eq:Rie}. 
Then we get $R \geq -1$ or, equivalently, $E_{cc}\leq - E_{bc}$, which means that a repulsive continuum self-interaction cannot be stronger than the bare-continuum interaction.
For an attractive continuum self-interaction, we consider the model ``bcc''.
First, because in this case we have $E_C>m>E_{\text{th}}$, a CDD zero can be observed as a zero of the phase shift.
Then from \cref{eq:prR}, $R \geq 1$ gives
\begin{align}
    E_C \leq m + \frac{m-E_B}{2}\,.
\end{align}
Using $m = 2480 \,\text{MeV}$, we get $E_C \leq 2561\,$MeV. Using $m = 2406 \,\text{MeV}$, we get $E_C \leq 2450\,$MeV.
If future data excludes the presence of a CDD zero below these energies, the model ``bcc'' will support the conclusion that the bare-continuum interaction is the stronger force inside $D_{s0}^*(2317)$, and thus the bare $(c\bar{s})$ state is crucial for the generation of $D_{s0}^*(2317)$.
Of course, this criterion is model dependent, and its uncertainty needs further exploration.
We also note that the $D_s\eta$ channel with threshold energy $E_{D_s\eta}=2516\,$MeV can also significantly contribute to the generation of $D_{s0}^*(2317)$. This also needs further exploration.

\section{Summary}\label{sec:sum}
In this work, we have studied several toy potential models including their inverse scattering problems.
We have also discussed how they can be applied to analyze real-world systems.

The model ``cc'' has the Hamiltonian \cref{eq:Hcc}.
The potential of this model takes a separable form, and hence is also called the separable potential model.
We first showed that this model can typically approximate the $S$-wave near-threshold physics of a system with a relative error of $\mathcal{O}\left(\beta^3/M_V^3\right)$, providing the absence of a deeply bound state. 
Here $\beta$ is the largest near-threshold momentum and $M_V$ is the typical scale of the potential.
Then, the inverse scattering representations of some quantities were derived, including the potential \cref{eq:f2pcc}, and the possible bound state's wave function \cref{eq:wfcc}.

The model ``bc'' has the Hamiltonian \cref{eq:Hbc}.
This model considers a single bare state, and only includes the interaction between the bare state and the continuum states.
This model can be used as an approximation that ignores the continuum states' self-interaction.
The derived inverse scattering representations include the bare mass \cref{eq:bmbc}, the potential \cref{eq:f2pbc}, the bare-state distribution \cref{eq:Pbc}, and the bare-state proportion $Z$ \cref{eq:Zbc} and the wavefunction \cref{eq:wfbc} of the possible bound state.
In the end, we also showed that a shallow bound state tends to have a small bare-state proportion $Z$, even though the bare state plays a crucial role in this model.

The model ``bcc'' has the Hamiltonian \cref{eq:Hbcc}.
This model is a combination of the first two models.
It can be used to study the correction of the model ``bc''.
A new feature compared to the model ``bc'' is the presence of a CDD zero, which is a zero of the on-shell $T$-matrix element.
The presence of the zero is recognized as an interplay between the bare-continuum interaction and continuum self-interaction.
The derived inverse scattering representations include those in model ``bc''.
%
We also introduced the potential ratio \cref{eq:prR} for a bound state, which quantifies the relative strength of the continuum self-interaction and the bare-continuum interaction when forming the bound state.

These models were then applied to study several real-world systems.
The deuteron was analyzed using the model ``cc''.
Though the prediction for the wave function of the deuteron coincides with that in Ref.~\cite{Li:2021cue}, the model ``cc'' is expected to have an uncertainty of $\mathcal{O}(b^{3}/m_\pi^{3}=4\%)$.
The $\rho$ meson was analyzed using both the model ``bc'' and the model ``bcc''.
The results have already been summarized in the last paragraph of \cref{sec:ppr}.
Finally, the potential ratio $R$ of $D_{s0}^*(2317)$ in the $DK$-$(c\bar{s})$ system was analyzed.
We first showed that model independently a repulsive $DK$-$DK$ interaction cannot be stronger than the $DK$-$(c\bar{s})$ interaction inside $D_{s0}^*(2317)$.
Then, for the attractive $DK$-$DK$ interaction, based on the model ``bcc'', we provided a criterion that the bare $(c\bar{s})$ state is crucial for the generation of $D_{s0}^*(2317)$ if future experiments exclude the presence of a CDD zero below $2561\,$MeV or $2450\,$MeV.
The two energies come from two different quark-model predictions for the bare $(c\bar{s})$ mass from Ref.~\cite{Godfrey:1985xj} and Ref.~\cite{Yang:2021tvc} respectively.
We emphasize that this criterion not only is model dependent, but also ignores the effect of the $D_s\eta$ channel.

\section*{Acknowledgements}
We would like to thank Feng-Kun Guo, Hao-Jie Jing and Jin-Yi Pang for their helpful comments and discussions. 
This work is supported by National Natural Science Foundation of China under Grant No. 12175239 and by the National Key R\&D Program of China under Contract No. 2020YFA0406400, and by the Key Research Program of the Chinese Academy of Sciences, Grant NO. XDPB15.  

\begin{appendix}

\section{Bloch-Horowitz theory}\label{app:BH}
Consider the separation of the Hilbert space: $1=P+Q$, where $P$ and $Q$ are projection operators of two subspaces.
The Schr\"{o}dinger equation $H \ket{\psi} = E \ket{\psi} $ becomes
\begin{align}
    H_{PP}\ket{\psi_P} + H_{PQ}\ket{\psi_Q} &=E\ket{\psi_P} \,,\label{eq:seBH1} \\
    H_{QP}\ket{\psi_P} + H_{QQ}\ket{\psi_Q} &=E\ket{\psi_Q} \,,\label{eq:seBH2}
\end{align}
where $H_{IJ}:=IHJ$ and $\ket{\psi_I}:=I\ket{\psi}$ with $I,J=P,Q$.
\cref{eq:seBH2} gives
\begin{align}
    \ket{\psi_Q} = \frac{1}{E-H_{QQ}}H_{QP}\ket{\psi_P} \,.
\end{align}
Then, substituting it back into \cref{eq:seBH1} gives an effective Schr\"{o}dinger equation:
\begin{align}\label{eq:ese}
    H_\text{eff}(E)\ket{\psi_P}=E\ket{\psi_P}\,,
\end{align}
where $H_\text{eff}(E)$ is the energy-dependent Bloch-Horowitz effective Hamiltonian \cite{Bloch:1958determination,Luu:2005qr}, defined as
\begin{align}
    H_\text{eff}(E):=H_{PP}+H_{PQ}\frac{1}{E-H_{QQ}}H_{QP}\,.
\end{align}

If the original state is normalized as $\braket{\psi|\psi}=1$, its $P$-space projection will be normalized according to \cite{Lepage:1997How}
\begin{align}
    1 = \bra{\psi_P}\left( 1 + H_{PQ}\frac{1}
    {(E-H_{QQ})^2}H_{QP} \right) \ket{\psi_P} = \bra{\psi_P}\left( 1 - \frac{\partial\,H_\text{eff}(E) }{\partial\,E}\right) \ket{\psi_P} \,.
\end{align}

It is also convenient to formalize the above discussion when the Hamiltonian can be separated as $H=H_0+V$.
Then, we have
\begin{align}
    H_\text{eff}(E):=H_{0;PP}+V_{PP}+V_{PQ}\frac{1}{E-H_{0;QQ}-V_{QQ}}V_{QP} \equiv H_{0;PP} + V_\text{eff}(E) \,.
\end{align}

\section{Approximating $S$-wave near-threshold physics with a separable potential}\label{app:assep}
We define the $Q$ space in the Bloch-Horowitz theory discussed in \cref{app:BH} to be all the states with momentum $q>Q$.
Then for the real-world potential $V^{\text{(rw)}}(p,k)$, one gets an effective potential in the remaining $P$ space:
\begin{align}\label{eq:veff}
    V_{\text{eff}}(p,k;E)&=V(p,k)+ \braket{p|VQ\frac{1}{E-H_{QQ}}QV|k} \nonumber\\
    &= \left[V(p,k) + \braket{p|VQ\frac{1}{E_0-H_{QQ}}QV|k}\right] + \braket{p|VQ\frac{-(E-E_0)}{(E-H_{QQ})(E_0-H_{QQ})}QV|k} \nonumber\\
    &\equiv [V_{e1}(p,k) + V_{e2}(p,k)] +  V_f(p,k;E)  \equiv  V_e(p,k) + V_f(p,k;E) \,,
\end{align}
where we omit the superscript ``(rw)'' to simplify the notation, and $E_0$ can be chosen as any fixed energy close to $E_{\text{th}}$.

On the other hand, the same applies to the separable potential
\begin{align}
    V^{(f)}(p,k) = f(p)f(k) \,,
\end{align}
which gives
\begin{align}
    V^{(f)}_e(p,k)&=a\,f(p)f(k) \,, \nonumber\\
    V^{(f)}_f(p,k;E)&=b(E)f(p)f(k) \,.
\end{align}
where
\begin{align}
    a &= 1+\braket{f_Q|\frac{1}{E_0-H^{(f)}_{QQ}}|f_Q}\,, \nonumber\\ 
    b(E) &= \braket{f_Q|\frac{-(E-E_0)}{(E-H^{(f)}_{QQ})(E_0-H^{(f)}_{QQ})}|f_Q} \,,
\end{align}
and
\begin{align}
    \ket{f_Q}=Q\ket{f}=\int_Q^\infty \frac{q^2dq}{(2\pi)^3} f(q)\ket{q} \,.
\end{align}

Here we want $V^{(f)}_e(p,k)$ to approximate $V_e(p,k)$ up to second order in momenta, and $V^{(f)}_e(p,k)$ to approximate $V_e(p,k)$ up to first order in momenta.
Considering the expansion $f(p)=f_0+f_2 p^2+\mathcal{O}(p^4)$, one gets
\begin{align}\label{eq:vfef}
    V^{(f)}_e(p,k)&=af_0^2 + a f_0 f_2(p^2+k^2) + \mathcal{O}\left[p^4,k^4,p^2k^2\right] \,, \nonumber\\
    V^{(f)}_f(p,k;E)&=b(E) f_0^2  + \mathcal{O}\left[p^2(E-E_0),k^2(E-E_0)\right] \,,
\end{align}
Then we need
\begin{align}
    af_0^2 &= V_e(0,0) \,, \nonumber\\
    af_0f_2 &= \frac{\partial }{\partial p^2} V_e(p,0) \Big{|}_{p=0} \,, \nonumber\\
    b(E) f_0^2 &= V_f(0,0,E) \,.
\end{align}
The undetermined quantities include $f_0$, $f_1$, and all the $f^2(q>Q)$.
The first two determine the low-momentum behavior of $f(p)$, while the last one determines the high-momentum behavior, so they can be determined independently.
The last one $f^2(q>Q)$ indicates that we have infinite degrees of freedom. On the other hand, the last equation above is $E$ dependent, and can be viewed as infinite equations.
One can expand the equation as a power series of $(E-E_0)$, and retaining a finite number of terms can ensure the existence of a solution.
Then one should add $\mathcal{O}\left[(E-E_0)^n\right]$ to $V^{(f)}_f(p,k;E)$ in \cref{eq:vfef}, where $n$ is the cutoff of the power-series expansion of $(E-E_0)$, and is assumed to be large enough so that the resulting uncertainty can be ignored in the following error analysis.

Now we analyze the error from this approximation.
We label the highest momentum of the states that we want as $\beta$.
According to the effective Schr\"{o}dinger equation \cref{eq:ese}, all of the momenta $p,k$ in the $P$ space will be coupled together, and the energy $E$ is set to be definite. Therefore, we require $V_\text{eff}(p\leq Q,k\leq Q;E\leq\frac{\beta^2}{2\mu})$.
We also label the typical momentum scale of $V(p,k)$ to be $M_V$.
As discussed around \cref{eq:yukawa}, we should have $V(p,k)=\mathcal{O}(M_V^{-2})$.
We will consider the following ordering:
\begin{align}
    \beta < Q < M_V < \mu \,,
\end{align}
where the last inequality ensures the nonrelativistic approximation $h_{M_V}\approx \frac{M_V^2}{2\mu}$.

$H_{QQ}$ is the Hamiltonian in the $Q$ space. We first assume that there are no bound states. Then the eigenstates of $H_{QQ}$ can be chosen as scattering ``in'' states labeled as $\ket{q_Q^+}$ with eigenvalue $h_{q_Q}$, and $q_Q$ starts from $Q$.
So we have
\begin{align}
    V_{e2}(p,k)&=\int_Q^\infty \frac{q_Q^2dq_Q}{(2\pi)^3} 
    \frac{\braket{p|V|q_Q^+}\braket{q_Q^+|V|k}}{E_0-h_{q_Q}} \,,\nonumber\\
    V_{f}(p,k;f)&=-(E-E_0) \int_Q^\infty \frac{q_Q^2dq_Q}{(2\pi)^3} 
    \frac{\braket{p|V|q_Q^+}\braket{q_Q^+|V|k}}{(E-h_{q_Q})(E_0-h_{q_Q})} \,.
\end{align}
Because the states $\ket{q_Q^+}$ and the free states $\ket{q>Q}$ span the same Hilbert space, and share the same normalization convention, and recalling $\braket{p|V|k}=\mathcal{O}(M_V^{-2})$, we expect also $\braket{p|V|q_Q^+}=\mathcal{O}(M_V^{-2})$.
The effective potential now scales as follows:
\begin{align}
    V_{e1}(p,k) &= V(p,k) = \mathcal{O}\left[\frac{1}{m_V^2}\right] \,,\nonumber\\
    V_{e2}(p,k) &= \mathcal{O}\left[\frac{1}{m_V^2}\left( \frac{\mu Q}{M_V^2} + \frac{\mu }{M_V} \right)\right] = \mathcal{O}\left[\frac{1}{m_V^2} \frac{\mu }{M_V} \right] \,, \nonumber\\
    V_f(p,k;E) &= \mathcal{O}\left[\frac{1}{m_V^2}\left( \frac{\mu\beta^2} {QM_V^2} + \frac{\mu\beta^2 }{M_V^3} \right)\right] =\mathcal{O}\left[\frac{1}{m_V^2} \frac{\mu\beta^2}{QM_V^2} \right] \,.
\end{align}
Here, for $V_{e2}$ and $V_f$, we consider both $q_Q\sim Q$ and $q_Q\sim M_V$ regions in the $Q$-space integration. The $q_Q\sim M_V$ region dominates $V_{e2}$, while the $q_Q\sim Q$ region dominates $V_f$.
We also note that it is the $q_Q\sim M_V$ region of $V_{e2}$ that dominates the whole $V_{\text{eff}}$.
When $H_{QQ}$ forms a shallow bound state, we expect that the above power counting still holds. When it forms a deeply bound state, however, the factor $\frac{1}{E-H_{QQ}}$ can ruin our power-counting analysis.
The presence of a deeply bound state of $H_{QQ}$ indicates that $V_{QQ}$ is a strong attractive interaction. Typically we would expect the attraction to get stronger for lower momentum, as in the case of one-boson-exchange potential. Therefore, $H_{QQ}$ cannot form a deeply bound state if $H$ does not form one.

Considering that $V_e$ has been approximated up to the second order of momenta, and $V_f$ only to the first order, the relative error should then scale as
\begin{align}
    \frac{\Delta V_e + \Delta V_f}{V_{\text{eff}}} &= \mathcal{O}\left[ \frac{Q^4}{M_V^4} + \frac{Q^2}{M_V^2}\frac{\beta^2}{QM_V} \right] \,.
\end{align}
Recalling that $Q$ is an artificial scale, we can set $Q \sim \beta$ to get
\begin{align}
    \frac{\Delta V_{\text{eff}}}{V_{\text{eff}}} = \mathcal{O}\left( \frac{\beta^3}{M_V^3} \right)  \,,
\end{align}
which holds for all $V_\text{eff}(p\leq Q,k\leq Q;E\leq\frac{\beta^2}{2\mu})$.
We also note that we cannot set $Q=\beta$, because the factor $\frac{1}{E-h_{h_{qQ}}}$ can touch its singularity and ruin our power-counting analysis. However, we can set $Q$ a bit higher than $\beta$ so that $\frac{1}{E-h_{h_{qQ}}}$ is still of order $\mathcal{O}(\mu/\beta^2)$.

\end{appendix}

\bibliography{refs}

\end{document}